\documentclass[english,aps,pra, twocolumn, floatfix, showpacs,  notitlepage]{revtex4-1}
\usepackage{lipsum}
\usepackage{amsmath}
\usepackage{amsfonts}
\usepackage{graphicx}
\usepackage{color}
\usepackage{hyperref}
\usepackage{epstopdf}
\begin{document}
 \title{Conductivity of lattice bosons at high temperatures}
 \author{Ivana Vasi\'c}
\author{Jak\v sa Vu\v ci\v cevi\'c}
\affiliation{Institute of Physics Belgrade, University of Belgrade, Pregrevica 118, 11080 Belgrade, Serbia
}
 \begin{abstract}

Quantum simulations are quickly becoming an indispensable tool for studying particle transport in correlated lattice models.
One of the central topics in the study of transport is the bad-metal behavior, characterized by the direct current (dc) resistivity linear in temperature.
In the fermionic Hubbard model, optical conductivity has been studied extensively, and a recent optical lattice experiment has demonstrated bad metal behavior in qualitative agreement with theory. Far less is known about transport in the bosonic Hubbard model.
We investigate the conductivity in the Bose-Hubbard model, and focus on the regime of strong interactions and high-temperatures.
We use numerically exact calculations for small lattice sizes.
At weak tunneling, we find multiple peaks in the optical conductivity that stem from the Hubbard bands present in the many-body spectrum.
This feature slowly washes out as the tunneling rate gets stronger.
At high temperature, we identify a regime of $T$-linear resistivity, as expected.
When the interactions are very strong, the leading inverse-temperature coefficient in conductivity is proportional to the tunneling amplitude.
As the tunneling becomes stronger, this dependence takes quadratic form.
At very strong coupling and half filling, we identify a separate linear resistivity regime at lower temperature, corresponding to the hard-core boson regime.
Additionally, we unexpectedly observe that at half filling, in a big part of the phase diagram, conductivity is an increasing function of the coupling constant before it saturates at the hard-core-boson result.
We explain this feature based on the analysis of the many-body energy spectrum and the contributions to conductivity of individual eigenstates of the system.
\end{abstract}

 \maketitle

\section{Introduction}

Cold atoms in optical lattices have provided a clean and tunable realization of the Hubbard model \cite{RevModPhys.80.885}. The focus of early experiments was on studying 
phase transitions within the model, but various aspects of nonequilibrium dynamics
have also been explored in this setup.
In particular, a lot of effort is invested in performing transport measurements with cold atoms \cite{2015NatPh11998C, Krinner_2017,Brown2019}.
Transport measurements in optical lattices are of great interest as they allow to isolate the effects of strong correlations from the effects of phonons and disorder, in a way that is not possible in real materials.

Particular attention is paid to linear-in-temperature resistivity, which is believed to be related to the superconducting phase and/or quantum critical points in the cuprates and more general strongly correlated systems\cite{Grigera2001,Cooper2009,Cao2018,Legros2018,Licciardello2019,Cha2020}.
This phenomenon has been studied theoretically in different versions of the Fermi-Hubbard model and in different parameter regimes\cite{Deng2013, Vucicevic2015,Perepelitsky,Vucicevic2019,Vranic2020,Kiely2021,Vucicevic2023}.
The onset of resistivity linear in temperature has been addressed in more general terms from the theoretical side in Refs.~\cite{HuseFermions,Herman2019, Patel_2022}.
In experiment with fermionic cold atoms in optical lattices, the $T$-linear resistivity has also been observed to span a large range of temperature, in qualitative agreement with theory\cite{Brown2019}.
However, the transport in bosonic lattice models has been less studied, from both theoretical and experimental perspectives.

Bosonic transport in the strongly interacting regime of the Bose-Hubbard model has been addressed in a cold-atom setup by investigating expansion dynamics induced by a harmonic-trap removal \cite{PhysRevB.88.235117, PhysRevLett.110.205301} and by studying center-of-mass oscillations induced by a trap displacement \cite{PhysRevA.76.051603, 2019PSSBR.25600752D}. However these studies do not focus on optical conductivity. Optical conductivity of bosons at zero and low temperature has been calculated in early papers \cite{FisherFisher, Cha, Kampf}. Conductivity of two-dimensional hard-core bosons has been addressed in Refs.~\cite{LindnerBosons, Bhattacharyya} and a large temperature range with linearly increasing resistivity has been found. Conductivity of strongly correlated bosons in optical lattices in a synthetic magnetic field was obtained in Ref.~\cite{Sajna}. The regime of resistivity linear temperature has been recently investigated for the Bose-Hubbard model at weak coupling \cite{Rizzatti}.
Dynamical response within the scaling regime of the quantum critical point and universal conductivity at the quantum phase transition have been investigated theoretically in an attempt to establish a clear connection with ADS-CFT mapping \cite{PhysRevLett.112.030402, PhysRevLett.118.056601}.
In addition to cold atom studies, bosonic transport properties have been studied in the context of an emergent Bose liquid \cite{2021arXiv211205747Z, 2021PNAS11800545H, Yue_2023}. Transport properties of a nanopatterned YBCO film arrays have been analyzed in terms of bosonic strange metal featuring resistivity linear in temperature down to low temperatures \cite{2022Natur.601..205Y}.

In this paper, we study conductivity in the Bose-Hubbard model\cite{FisherFisher, RevModPhys.80.885} as relevant for optical lattice experiments with bosonic atoms.
We consider strongly-interacting regime and focus on high temperatures, away from any ordering instabilities.
We consider small lattice sizes of up to $4\times 4$ lattice sites, 
and employ averaging over twisted boundary conditions to lessen the finite-size effects.
We control our results
by comparing different lattice sizes, as well as by checking sum rules to make sure that charge stiffness is negligible.
We also use both the canonical and grand canonical ensemble, and compare results.
To solve the model, we use exact diagonalization and finite-temperature Lanczos method\cite{prelovvsek2013strongly}.

We compute and analyze the probability distribution of the eigenenergies (the many-body density of states), the spectral function, the optical and the dc conductivity, as well as some thermodynamic quantities.
Where applicable, we compare results to the hard-core limit, the classical limit as well as to the results obtained by the bubble-diagram approximation.
Our results show several expected features.
First, the many-body density of states, spectral function and the optical conductivity, all simultaneously develop gaps and the corresponding Hubbard bands as the coupling is increased. Next, we clearly identify the linear dc resistivity regime at high temperature, and find the connection of the slope with the tunneling amplitude, along the lines of Ref.~\cite{Patel_2022}.
Furthermore, at lower temperatures and high coupling, we identify a separate linear resistivity regime corresponding to hard-core-like behavior, which can be expected on general grounds.
We also find some unexpected features: we observe non-monotonic behavior in the dc conductivity as a function of the coupling constant, which we map out throughout the phase diagram.

The paper is organized as follows: In Sec.~\ref{sec:modelandmethods}, we briefly describe our method of choice. In Sec.~\ref{sec:results} we present our results: in subsection \ref{sec:espectrum} we address the many-body density of states and in subsection \ref{sec:th} we show some thermodynamic properties of the Bose-Hubbard model in the high-temperature regime. In subsection \ref{sec:opticalconductivityresults} we calculate optical conductivity for a finite interaction strength  and in subsection \ref{sec:dclimit} we investigate the dependence of the direct-current conductivity on microscopic parameters of the Bose-Hubbard model and temperature. In subsection \ref{sec:Kubof} we explain observed features using an analysis of the Kubo formula from Ref.~\cite{Patel_2022}. Then in subsection \ref{sec:hardcorebosons} we compare our results for half filling with the results for hard-core bosons \cite{LindnerBosons}. We discuss finite-size effects in subsection \ref{sec:fse} and compare results obtained within the canonical ensemble with the results obtained within the grand-canonical ensemble in subsection \ref{sec:gce}. In subsection \ref{sec:bubble} we compare our results obtained for small lattices with the result of often used bubble-diagram approximation. Finally, we summarize our findings in Sec.~\ref{sec:conclusions}.

 \section{Model and Methods}
 \label{sec:modelandmethods}

Cold bosonic atoms in optical lattices are realistically described by the Bose-Hubbard model \cite{RevModPhys.80.885}
\begin{equation}
 H = -J \sum_{\langle ij\rangle}\left(b_i^{\dagger}b_j + \mathrm{H. c.}~\right)+\frac{U}{2}\sum_i n_i \left(n_i-1\right),
 \label{eq:bh}
\end{equation}
where $J$ is the tunneling amplitude between nearest-neighbor sites of a square lattice and $U$ is the on-site density-density interaction.
 Unless stated differently, our units are set by the choice $U = 1$.
Throughout the paper, we set lattice constant $a = 1$, $\hbar = 1$, $k_\mathrm{B} = 1$.
We also assume that effective charge of particles is $q=1$.

The quantitative finite-temperature phase diagram of the model on a square lattice was obtained in Ref.~\cite{Capogrosso-Sansone}. At integer filling, a quantum phase transition between a Mott insulator
state and a superfluid  is found (in particular, for filling $n = 1$ boson per site, the transition occurs at $(J/U)_c \approx  0.0597$.
At finite temperature a BKT transition describes the loss of superfluidity \cite{PhysRevLett.87.270402}.
In this paper we work in the high-temperature regime where we expect only normal (non-condensed) state and short-range correlations.

We consider the retarded current-current correlation function at finite temperature $T=1/\beta$
\begin{equation}
 C_{xx}(t) =  -i \theta(t)\langle [J_x(t), J_x] \rangle_{\beta},
 \label{eq:ct}
\end{equation}
where the current operator is given by
\begin{equation}
 J_x = -i J\sum_{\langle ij\rangle_x} \left(b_i^{\dagger}b_{j}-\mathrm{H.\,c.}~\right),
\end{equation}
with the summation only over nearest-neighbors along $x$ direction. The partition function is $Z(\beta) = \mathrm{Tr} \left(e^{-\beta  H}\right)$ and averaging is performed as
\begin{equation}
 \langle X \rangle_{\beta} =  \frac{1}{Z(\beta)} \mathrm{Tr} \left(e^{-\beta  H} X\right).
\end{equation}
In the linear response regime, the conductivity is given by the Kubo formula\cite{Mahan, Scalapino}
\begin{equation}
 \sigma_{xx}(\omega) = \frac{i}{\omega} \left(\langle -E_{\mathrm{kin}}^x\rangle_{\beta} + C_{xx}(\omega)\right),
 \label{eq:sigmaxx}
\end{equation}
where $C_{xx}(\omega) =  \int_{-\infty}^{\infty} dt e^{i \omega t} C_{xx}(t)$ and $\langle E_{\mathrm{kin}}^x\rangle_{\beta}$ is the average kinetic energy along $x$ direction. 
Note that for the real conductivity from Eq.~(\ref{eq:sigmaxx}), we need imaginary part of the correlation function $\textrm{Im}\, C_{xx}(\omega)$. The direct-current (DC) conductivity is obtained as $\sigma_{\mathrm{DC}}=\mathrm{Re}\,\sigma(\omega \rightarrow 0)$. 

In the following we rely on numerically exact approaches for small lattice sizes.
We use exact diagonalization to obtain the eigenenergies $E_n$ and eigenstates $|n\rangle$ of the Hamiltonian $H$ and calculate
\begin{eqnarray}
\text{Re}\, \sigma_{xx}(\omega)=\frac{\pi}{Z(\beta)\omega}\sum_{n, m} &&|\langle n|J_x|m\rangle|^2\left(e^{-\beta E_n}-e^{-\beta E_m}\right)\nonumber\\&\times&\delta\left(\omega  + E_n - E_m\right).
\label{eq:cxx}
 \end{eqnarray}
For larger lattice sizes, we apply the finite-temperature Lanczos method \cite{prelovvsek2013strongly}. We employ averaging over twisted boundary conditions to lessen the finite-size effects \cite{PhysRevB.44.9562, 1992ZPhyB..86..359G}.

 Due to the expected onset of charge stiffness in finite systems and, more generally, longer range correlations at low temperature, we expect our method to be limited to the regime of high temperature and relatively strong interactions (corresponding to a smaller ratio $J/U$).
For a crosscheck of the validity of our numerical approaches, we rely on  the comparison of different lattice sizes (subsection \ref{sec:fse}) and the sum rule
\begin{equation}
 \int _0^{\infty}\!d\omega\,\text{Re}\,\sigma_{xx}(\omega) = \frac{\pi}{2} \langle -E_{\mathrm{kin}}^x\rangle_{\beta}.
 \label{eq:sumrule}
\end{equation}
Indeed, we find that the sum rule is satisfied at temperatures $T/U\geq1$ and tunneling $J/U\leq0.2$. In this regime we find that the results no longer significantly change with increasing lattice size (subsection \ref{sec:fse}), thus our results are expected to be reasonably representative of the thermodynamic limit.

Finally, we control the results with respect to the choice of the statistical ensemble.
A comparison of the results obtained with the grand-canonical and canonical ensembles is given in Sec.~\ref{sec:gce}.
We do not find substantial difference in the results between the two statistical ensembles, thus we choose to work in the canonical ensemble, as it allows us to consider larger lattices. Unless stated differently, the results presented in the paper are for the canonical ensemble.

\section{Results}
\label{sec:results}

\subsection{Eigenstate spectrum}
\label{sec:espectrum}

\begin{figure}[!t]\includegraphics[width=0.5\textwidth]{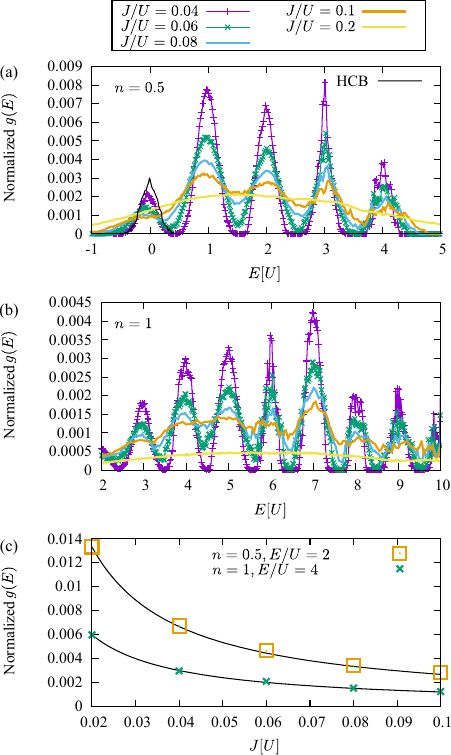}
\caption{The normalized many-body density of states defined in Eq.~(\ref{eq:dos}) for (a) $n = 0.5$,  $L_x = 4, L_y = 3, N_p = 6$ and (b) $n = 1$, $L_x = 3, L_y = 3, N_p = 9$. The black solid line in (a) gives the corresponding density of states for the case of hard-core bosons. We use $100$ random realizations of the twisted boundary conditions. In (c) we plot the height of the peak at energies $E = 2 U$ (for half filling) and $E = 4 U$ (for unit filling). The black solid lines give the fitted $(J/U)^{-1}$ dependence.}
\label{fig:Figdos}
\end{figure}
We first consider the many-body density of states of the model (\ref{eq:bh}) defined by
\begin{equation}
  g(E) = \sum_n\delta(E-E_n).
  \label{eq:dos}
\end{equation}
We show our numerical results for fillings $n=1/2$ and $n=1$ in Fig.~\ref{fig:Figdos}(a).

The classical limit $J=0$ is simple to understand.
The many-body spectrum features energies $E_n^{J=0} = n U, n = 0, 1,\ldots$ with huge degeneracies. The Hilbert space dimension is $\mathrm{dim}\,\mathcal{H} = \binom{L+N_p-1}{ N_p}$, where $L$ is the number of lattice sites, and $N_p$ is the number of particles. For $N_p<L$, there are $\binom{L} {N_p}$ states with zero energy, the  $\binom{L} {1} \binom{L-1}{N_p-1}$ states (a single-site occupied with two bosons) with energy $E_n=U$, and so on.  The number of different energy levels is set by the system size that we consider. As the ratio $J/U$ increases from zero, these macroscopic degeneracies are slowly resolved, and the bands obtain a finite width. This is precisely what we observe in the numerical data, as shown in Fig.~\ref{fig:Figdos}(a), but we find that separate many-body bands do persist up to a finite value of $J/U \approx 0.05$, regardless of filling.
Around this value, the peaks in $g(E)$ (found around $E = n U$) have heights roughly proportional to $1/J$, see Fig.~\ref{fig:Figdos}(c). As $J$ increases further, the density of states turns into a wide, bell-shaped curve, as shown in Fig.~\ref{fig:Figdos}(b) for unit filling at $J/U = 0.2$.
This particular feature of the spectrum has been considered for a related one-dimensional model in Ref.~\cite{KollathJStat}.
We also observe that for $n=1/2$, the lowest many-body band can be reasonably approximated by hard-core bosons.

\subsection{Thermodynamic properties}
\label{sec:th}

\begin{figure}[!t]\includegraphics[width=0.5\textwidth]{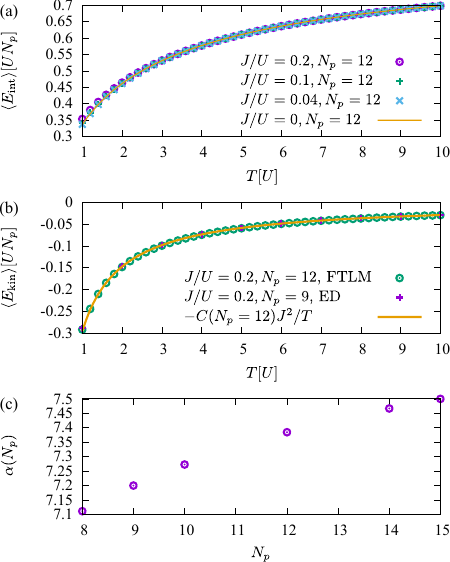}
\caption{(a) Interaction energy, Eq.~(\ref{eq:int}),  as a function of temperature. (b) Kinetic energy, Eq.~(\ref{eq:kin}), as a function of temperature. (c) The coefficient $ \alpha(N_p)$ from Eq.~(\ref{eq:kin}) as a function of system size. System size $L_x = 4, L_y = 3$, number of particles $N_p = 12$.}
\label{fig:Figenergy}
\end{figure}

We investigate thermodynamical properties for $T/U\geq1$, as this regime has not been discussed in much detail in literature. Interaction energy 
\begin{equation}
\langle E_{\text{int}}\rangle_{\beta}=\frac{U}{2}\langle  \sum_i n_i (n_i - 1)\rangle_{\beta}
\label{eq:int}
\end{equation}
is very weakly dependent on the tunneling $J/U$ as shown in Fig.~\ref{fig:Figenergy}(a), for the tunneling range that we consider $0\geq J/U\leq0.2$. The average value of the kinetic energy 
\begin{equation}
 \langle E_{\text{kin}}\rangle_{\beta}=\langle-J\sum_{\langle ij\rangle} b_i^{\dagger}b_j + \mathrm{H.\,c.}~\rangle_{\beta}
\end{equation}
can be reasonably approximated by the leading order in the high-temperature expansion
\begin{eqnarray}
 \langle E_{\mathrm{kin}} \rangle_{\beta}&\approx&\frac{\mathrm{Tr}(1-\beta H)E_{\text{kin}}}{\mathrm{Tr}(1-\beta H)}\\ &\approx& -\beta\frac{ \mathrm{Tr}E_{\text{kin}}^2}{\text{dim} \mathcal{H}}\\&=&-\alpha(N_p) N_p J^2/T,\\  \alpha(N_p)& =& \frac{1}{\text{dim} \mathcal{H}}\sum_{<ij>} \text{Tr}n_i (n_j+1)
 \label{eq:kin}
\end{eqnarray}
where $N_p$ is the number of particles in the system and $\text{dim} \mathcal{H}$ is the dimension of the Hilbert space. The coefficient $\alpha(N_p)$ is size dependent, see Fig.~\ref{fig:Figenergy}(c).  

\subsection{Optical conductivity}
\label{sec:opticalconductivityresults}

\begin{figure}[!t]
\includegraphics[width=0.5\textwidth]{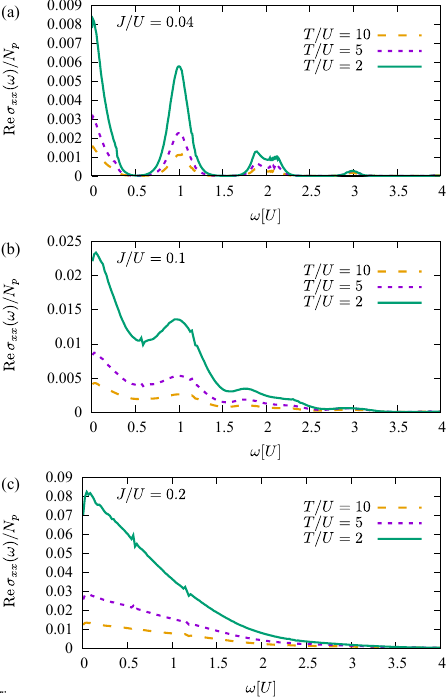}
\caption{Optical conductivity $\mathrm{Re}\,\sigma_{xx}(\omega)$ as a function of frequency. Parameters: (a) $J/U = 0.04$, (b) $J/U = 0.1$, (c) $J/U = 0.2$, $L_x \times L_y = 4 \times 3, N_p = 6$.  }
\label{fig:Figsigma}
\end{figure}
\begin{figure}[!h]\includegraphics[width=0.5\textwidth]{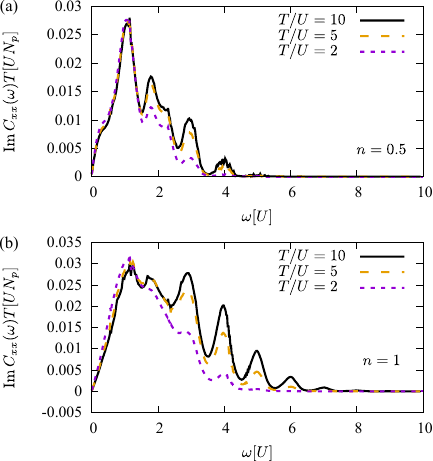}
\caption{Imaginary part of the current-current correlation function $\textrm{Im}\,C_{xx}(\omega)$ multiplied by temperature $T$ for a better comparison as a function of frequency $\omega$.  Parameters: (a)$N_p = 6$, (b) $N_p = 12$ and $J/U = 0.1, L_x \times L_y = 4 \times 3$.}
\label{fig:Figc}
\end{figure}

We present numerical results for the optical conductivity in Fig.~\ref{fig:Figsigma} for bosons with $J/U = 0.04, 0.1, 0.2$ and temperatures $T/U = 2, 5, 10$. We check that within this range of physical parameters the sum rule given in Eq.~(\ref{eq:sumrule}) is satisfied with accuracy better than one percent. For the weak tunneling amplitudes $J/U = 0.04$ we observe that the conductivity exhibits multiple peaks at $\omega\sim n U, n = 1,2, 3, \ldots$ that stem from the energy bands of the Hubbard model, as shown in Fig.~\ref{fig:Figdos}.  As temperature is lowered, the higher-energy peaks become smaller relative to the low-energy peaks, which is clearly expected. However, in absolute terms, the optical conductivity gets smaller with increasing temperature, at all frequencies. As the tunneling gets stronger the peaks merge, but the  multi-peak structure is still visible at intermediate hopping $J/U = 0.1$. Finally for $J/U = 0.2$ the conductivity takes a simpler form, see Fig.~\ref{fig:Figsigma} (c). In Fig.~(\ref{fig:Figc}) we show that at low frequency, the current-current correlation function scales with temperature in a simple way, but the behavior is more complicated at higher frequencies.

\begin{figure}[!th]\includegraphics[width=0.5\textwidth]{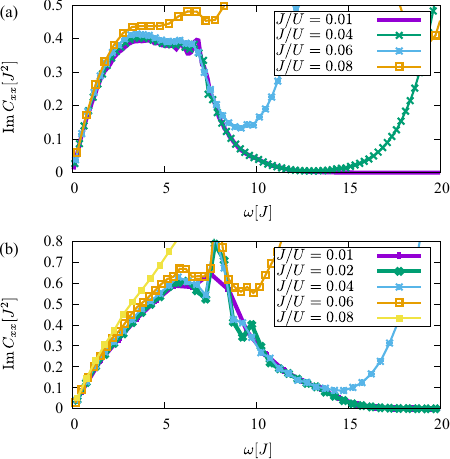}
\caption{Imaginary part of the current-current correlation function $\textrm{Im}\,C_{xx}(\omega)$ as a function of frequency, $T/U = 10$. Parameters: (a) $ L_x \times L_y = 4 \times 3, N_p = 6$, (b) $ L_x \times L_y = 3 \times 3, N_p = 9$.}
\label{fig:Figscaling}
\end{figure}

\subsection{DC limit and $T$-linear resistivity}
\label{sec:dclimit}

We now focus on the range of small $\omega$. We have seen that the strong interaction $U$ introduces gaps in the many-body spectrum that translate into peaks in the optical conductivity. In order to address the role of tunneling $J$ in more detail, we replot numerical data in Fig.~\ref{fig:Figscaling} by showing the current-current correlation $\text{Im} C_{xx}(\omega)/J^2$ as a function of $\omega/J$. The rescaling by $J^2$ is motivated by the basic definition of the current-current correlation function from Eq.~(\ref{eq:ct}). We find that for up to $J/U\approx 0.08$ numerical data overlap near $\omega=0$ and are very weakly dependent on $J/U$.  We further analyze and explain this feature in  Sec.~\ref{sec:Kubof}.

In order to extract the DC conductivity $\sigma_{\text{DC}}$
from the current-current correlation function, we consider small but finite $\omega$ and perform a linear fit $\lim_{\omega\rightarrow0}\textrm{Im}\,C_{xx}(\omega)(T) \approx \sigma_{\text{DC}}(T) \times \omega$. For numerical purposes we perform the fitting in the range $\omega\in(0, J)$.

\begin{figure}[!t]\includegraphics[width=0.5\textwidth]{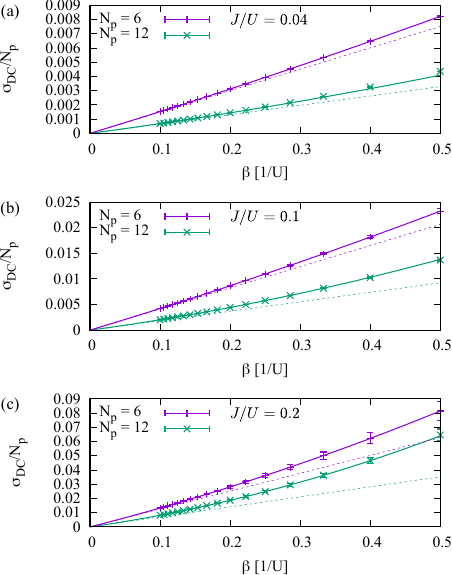}
\caption{DC conductivity $\sigma_{DC}$ as a function of inverse temperature $\beta$ for (a) $J/U = 0.04$, (b) $J/U = 0.1$, (c) $J/U = 0.2$. System size $L_x \times L_y = 4 \times 3$. The solid lines give fits according to Eq.~(\ref{eq:sigmabeta}), and dashed lines give the leading linear dependence in $\beta$.}
\label{fig:Figrho}
\end{figure}

In Fig.~\ref{fig:Figrho} we plot DC conductivity as a function of inverse temperature $\beta$. Overall, we find that the normalized conductivity $\sigma_{DC}/N_p$ decreases with filling from $n=1/2$ to $n=1$ in this strongly interacting regime. This is easily understood as, at integer fillings, the model is expected to have maximal resistivity (at integer filling, low temperature and strong enough coupling, the model is in the Mott insulating phase).
More importantly, we observe at high temperature a clear linear regime $\sigma_\mathrm{DC}\sim \beta$, i.e. $\rho_\mathrm{DC}\sim T$.

In general, it is expected that a finite-size system with a bounded  spectrum exhibits resistivity linear in temperature in the limit $T\rightarrow\infty$ (or $\sigma_{\mathrm{DC}}$ linear in inverse temperature $\beta$) \cite{HuseFermions, Patel_2022}. When working within the canonical ensemble in the strongly interacting limit, the upper bound of the Bose-Hubbard model is close to $N_p(N_p-1)U/2$, where $N_p$ is the number of particles, and all the bosons occupy the same site. Within the grand-canonical ensemble we control the average density by reducing the chemical potential $\mu$ and in this way we effectively limit the highest energy state.

\begin{figure}[!ht]\includegraphics[width=0.5\textwidth]{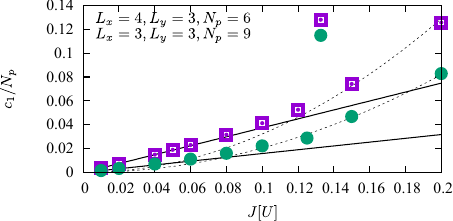}
\caption{The coefficient $c_1$ introduced in Eq.~(\ref{eq:sigmabeta}) versus tunneling amplitude $J/U$. The solid lines give linear fits and dashed line present quadratic functions.}
\label{fig:Figsummary}
\end{figure}

The key question is how far down in the temperature range resistivity linear in temperature persists. In particular, for hard-core bosons it has been shown that higher-order corrections become strong only at low temperatures in the vicinity of the BKT phase transition  \cite{LindnerBosons}.
To determine where quadratic corrections to $\sigma_{\text{DC}}\sim \beta$ start to play a role, we fit our result to
\begin{equation}
 \sigma_{\text{DC}}(\beta) \approx c_1 \beta + c_2 \beta^2
 \label{eq:sigmabeta}
\end{equation}
in the range $\beta U\in(0.1, 0.2)$.
We find that this approximation works well even for a wider range $\beta U\in(0.1, 0.5)$. For weak hopping $J/U\leq0.1$, in agreement with the observation from Fig.~\ref{fig:Figscaling}, we find that $c_1\sim J$, or
\begin{equation}
 \sigma_{\text{DC}}\sim J/T .
 \label{eq:JoverT}
\end{equation} 
A closely related result for the fermionic model has been derived using a high-temperature expansion in Ref.~\cite{Perepelitsky}.
For the stronger hopping it holds that $c_1\sim J^2$, as shown in Fig.~\ref{fig:Figsummary}. The sub-leading term $c_2$ becomes more prominent as the ratio $J/U$ gets stronger. 

\subsection{The analysis of the Kubo formula}
\label{sec:Kubof}

We explain the numerically identified features of the conductivity using the framework introduced in Ref.~\cite{Patel_2022}. Starting from the Kubo formula, Eq.~(\ref{eq:cxx}), we consider small enough $\omega\rightarrow 0$. Following the previous numerical analysis, in the following we use small but finite $\omega$.
We approximate the Dirac delta function from Eq.~(\ref{eq:cxx}) as $\delta(x)\approx\theta(\Delta\omega/2-|x|)/\Delta \omega $,  where $\theta$ is the Heaviside function, and the bin width $\Delta \omega$ is chosen to accommodate a reasonably large number of energy levels of our finite size system. By rewriting Eq.~(\ref{eq:cxx}) and taking the limit  $\omega\rightarrow0$ we obtain
\begin{widetext}
\begin{eqnarray}
\sigma_{DC}&=& \lim_{ \omega\rightarrow0}\frac{\pi\left(1-e^{-\beta \omega}\right)}{\omega Z(\beta) } 
\sum_{n,m}\delta(E_n-E_m-\omega)|\langle n|J_x|m\rangle|^2e^{-\beta E_n} \nonumber\\
&=& \frac{\pi\beta}{Z(\beta)}\lim_{ \omega\rightarrow0, \Delta \omega\rightarrow0}\frac{1}{  \Delta \omega} 
\sum_{n,m: |E_n-E_m-\omega|<\Delta\omega/2}|\langle n|J_x|m\rangle|^2e^{-\beta E_n} \nonumber\\
&=&\frac{\beta}{Z(\beta)} \sum_{n}lim_{ \omega\rightarrow0} f_n(\omega) e^{-\beta E_n},
\label{eq:app}
\end{eqnarray}
\end{widetext}
where
\begin{equation}
  f_n(\omega) = \lim_{ \omega\rightarrow0, \Delta \omega\rightarrow0}\frac{\pi}{\Delta \omega}\sum_{m: |E_n - E_m-\omega| <\Delta \omega/2} |\langle n|J_x|m\rangle|^2.
  \label{eq:cn}
\end{equation}
 
Using the high-temperature expansion in Eq.~(\ref{eq:app}), we find approximate results for the coefficients $c_1$ and $c_2$ introduced in Eq.~(\ref{eq:sigmabeta})
\begin{eqnarray}
 c_1&\approx& \langle f_n( \omega)\rangle_n\equiv\frac{1}{\text{dim} \mathcal{H}} \sum_n f_n( \omega),\label{eq:c1}\\
 c_2&\approx& \langle f_n( \omega)\rangle_n\langle E_n \rangle_n- \langle f_n( \omega ) E_n\rangle _n,
 \label{eq:c2}
 \end{eqnarray}
 where $\langle A_n\rangle_n\equiv \frac{1}{\text{dim} \mathcal{H}} \sum_n A_n$.
By a numerical inspection, we find that estimates for the coefficients $c_1$ and $c_2$ obtained in this way do match numerical data. 
Based on the approximation in Eq.~(\ref{eq:c1}) we infer that $c_1\sim J$ behavior is related to the presence of bands in the many-body spectrum. The bands' density of states is inversely proportional to $J$, see Fig.~\ref{fig:Figdos}, and consequently the number of available states within a frequency bin $\Delta \omega$ in Eq.~(\ref{eq:cn}) is inversely proportional to the tunneling $J$. For stronger $J$ a quadratic dependence $c_1\sim (J/U)^2$ appears  as the band structure is washed out, Fig.~\ref{fig:Figsummary}. 

\begin{figure}[!t]
\includegraphics[width=0.5\textwidth]{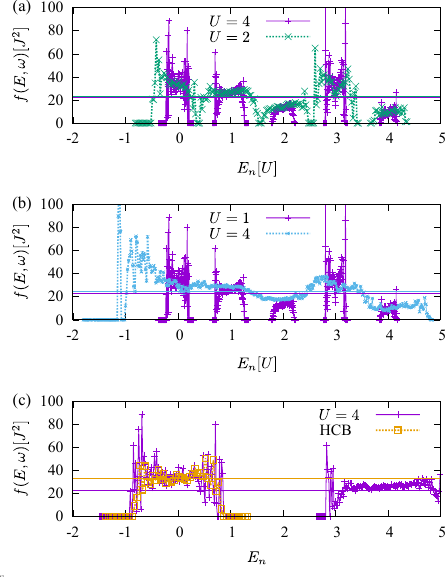}
\caption{The function $f(E, \omega)$ as defined in Eq.~(\ref{eq:cg}) for $J = 0.1, U = 4$ and (a) $J=0.1, U = 2$, (b) $J/U = 0.1, U = 1$, and (c) hard-core bosons with $J = 0.1$. Parameters: $ L_x \times L_y = 4 \times 3, N_p = 6$,  $\omega/J=0.35, \Delta \omega/U = 0.01$. The horizontal lines show averaged values of $f(E, \omega)$ over the full spectrum.}
\label{fig:FigTinv}
\end{figure}
Following Ref.~\cite{Patel_2022} we perform coarse-graining of the coefficients $f_n(\omega)$ from Eq.~(\ref{eq:cn}) 
\begin{equation}
 f(E, \omega)=\frac{1}{g(E)}\sum_n \delta\left(E_n-E\right) f_n(\omega),
 \label{eq:cg}
\end{equation}
where $g(E)$ is the many-body density of states, Eq.~(\ref{eq:dos}).  An almost flat line, an indicator of the invariance of $f(E, \omega)$ with energy $E$, ensures resistivity linear in temperature far down in $T$, as shown in Ref.~\cite{Patel_2022}. In Fig.~{\ref{fig:FigTinv}}(a) we present the coarse-grained coefficients $f(E, \omega)$ for $J = 0.1$ and $U = 2$ and $U = 4$. In both cases the many-body spectrum consists of the bands centered around $0, U, 2U, \ldots$. Within each band $f(E, \omega)$ is roughly constant and weakly dependent on $U$. As $U$ gets weaker, the gaps between bands are closed and the function $f(E, \omega)$ acquires more features, see  Fig.~{\ref{fig:FigTinv}}(b) for $U=1$. Because the function $f(E, \omega)$ deviates from an averaged flat line throughout the spectrum, at first glance the resistivity linear in temperature is found only at high-temperatures. This observation is in agreement with the data presented in Fig.~{\ref{fig:Figrho}} where the regime of linear resistivity is found roughly at $T\geq 5 U$ ($\beta\leq0.2$). However, in the next subsection we discuss another regime of linear resistivity found at lower temperatures when the hard-core description becomes relevant.

\subsection{Comparison with hard-core bosons}
\label{sec:hardcorebosons}

\begin{figure}[!t]
\includegraphics[width=0.5\textwidth]{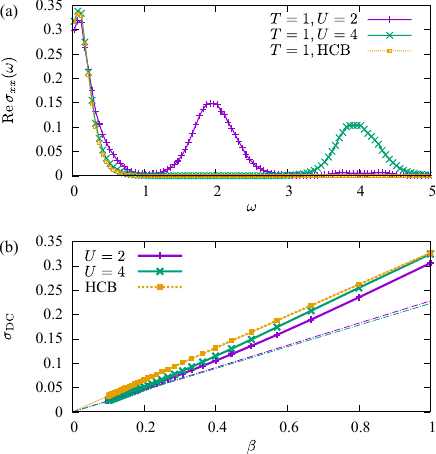}
\caption{(a) The conductivity $\mathrm{Re}\,\sigma_{xx}(\omega)$ vs.~frequency $\omega$ for half filling at $T = 1$ for hard-core bosons, $U = 2$ and $U = 4$. (b) The DC conductivity as a function of inverse temperature $\beta$. The dot-dashed lines give the leading-order high-temperature result $\sigma_{\mathrm{DC}} \approx c_1 \times \beta$ for $U = 4$ and $U = 2$. Parameters: $J = 0.1$,  $ L_x \times L_y = 4 \times 3, N_p = 6$.}
\label{fig:hcb}
\end{figure}

 Conductivity of hard-core bosons at half filling was investigated in Refs.~\cite{LindnerBosons}. It was found that a Gaussian function approximates well optical conductivity as a function of frequency. In order to reach the limit of hard-core bosons here we consider bosons at half filling ($n = 1/2$ bosons per lattice site). We keep fixed hopping rate $J = 0.1$ and change the value of local interaction $U$. In Fig.~\ref{fig:hcb}(a) we show that for the low enough ratio $T/U$, for example for $U = 4$ and $T = 1$, our results for the optical conductivity at half filling approach the result for hard-core bosons for $\omega/U < 1$ as expected. The contribution of higher Hubbard bands is absent in the hard-core model.

 The same applies to the DC conductivity as shown in Fig.~\ref{fig:hcb}(b). It was shown in Ref.~\cite{LindnerBosons} that the resistivity of hard-core bosons is linear in temperature down to very low temperatures close to the BKT transition. Our considerations from the previous subsection are in line with this conclusion, as we find the  hard-core bosons to exhibit nearly constant function $f(E, \omega)$, see Fig.~\ref{fig:FigTinv}(c), that closely corresponds to the function $f(E, \omega)$ of the lowest band of the full model. As expected, when the temperature is low enough with respect to $U$, only the lowest band of the full model is occupied and bosons can be described as hard-core particles. Consequently, the DC conductivity starts off as a linear function of the inverse temperature $\beta$, exhibits a  transitional behavior for a  range of intermediate $\beta$ values, and changes into a linear function corresponding to hard-core boson behavior at large $\beta$ (low $T$).

\begin{figure}[!t]
\includegraphics[width=0.5\textwidth]{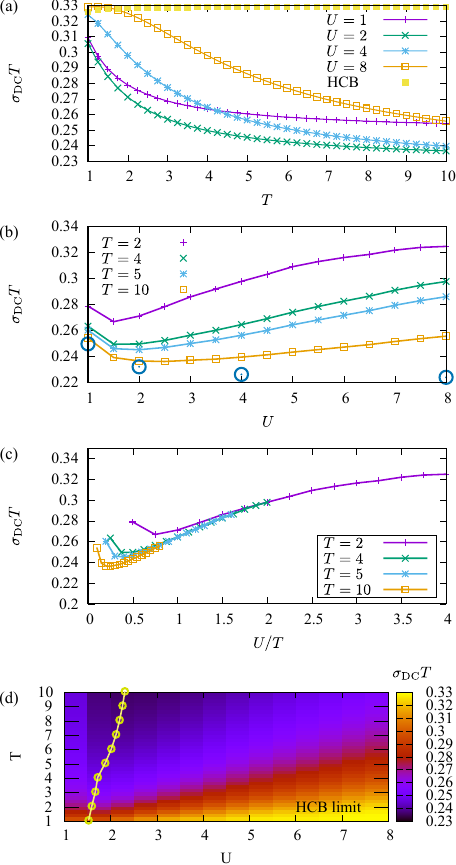}
\caption{The DC conductivity multiplied by temperature $T$ as a function of (a) temperature $T$, (b) interaction $U$, and (c) ratio $T/U$.  The large blue dots in (b) give the value of the coefficient $c_1$ introduced in Eq.~(\ref{eq:c1}) as a function of $U$. (d) The color map gives the DC conductivity multiplied by temperature $T$. The limit of hard-core bosons is reached in the bottom right corner of the plot. The yellow dots in in panel (d) give location of minima in $\sigma_{\mathrm{DC}} T$ as a function of $U$. The solid line is a guide to the eye. Parameters: $J = 0.1, L_x \times L_y = 4 \times 3, N_p = 6$.}
\label{fig:hcb2}
\end{figure}

 Now we investigate in more detail how the DC conductivity of hard-core bosons is reached by varying parameter $U$ of the full model and temperature $T$.  We present data for the DC conductivity as a function of temperature $T$ and interaction $U$ in Fig.~\ref{fig:hcb2}. As we plot the DC conductivity multiplied by temperature $T$, for hard-core bosons we find an almost perfect constant within the considered temperature range, see Fig.~\ref{fig:hcb2}(a). This constant is reached for $U=8$ at $T=2$ and for $U=4$ at $T=1$. What we find surprising is the $U$ dependence of the conductivity: we find $\sigma_{DC} T$ to exhibit a minimum at some finite $U$ before reaching the hard-core boson result, see Fig.~\ref{fig:hcb2}(b). 

 This behavior can be traced back to the result from Fig.~\ref{fig:FigTinv}(c) where we see that the averaged value of $f(E, \omega\rightarrow0)$ of hard-core bosons overestimates the result of the full model. We rewrite Eq.~(\ref{eq:app}) as
 \begin{equation}
  \sigma_{\mathrm{DC}}\, T =\frac{1}{\sum_{n} e^{-\beta E_n}} \sum_{n}\lim_{\omega\rightarrow0}f_n(\omega) e^{-\beta E_n},
  \label{eq:simp}
 \end{equation}
and consider the strongly interacting limit where the function $f_n(\omega\rightarrow0)$ can be approximated by a sum of rectangular functions centered around $0, U, 2U,\ldots$, as presented in Fig.~\ref{fig:FigTinv}(a). From Fig.~\ref{fig:FigTinv}(a) we observe that from the several lowest Hubbard bands, it is the lowest band that has the highest value of $f(E, \omega\rightarrow0)$.  From Eq.~(\ref{eq:simp}) we infer that at fixed temperature, as we increase the interaction strength $U$, both the nominator and the denominator of the last expression are reduced. Yet, the partition function $Z(\beta)$ decays faster, because each term $e^{-\beta E_n}\sim e^{-\beta n U}$ in the numerator is multiplied with a factor $f_n<1$ (and getting smaller with increasing $n$), while in the denominator it is multiplied by one. Therefore, the conductivity \emph{increases} with increasing U, and we reach the limit of hard-core-boson conductivity from below. This is a striking and highly counterintuitive observation. We have checked that this feature persists even when calculations are done in the grand canonical ensemble, and when the size of the lattice is changed, but it still might be an artifact of the finite size of the system.

\subsection{Finite-size effects}
\label{sec:fse}
\begin{figure}[!t]
\includegraphics[width=0.5\textwidth]{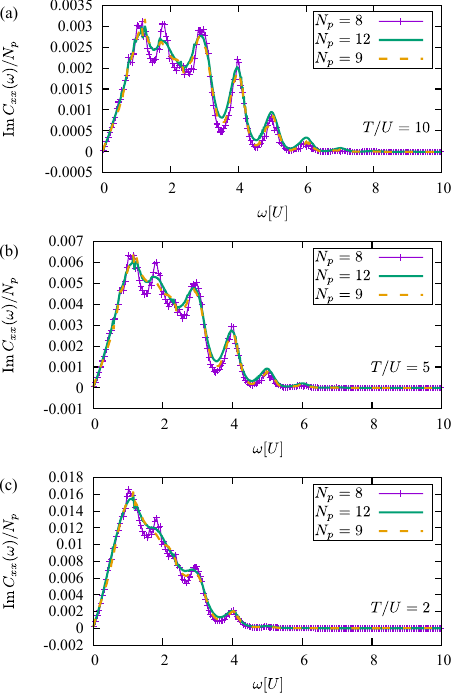}
\caption{The current-current correlation function vs.~frequency $\omega$ for unit filling and three different system sizes $L_x\times L_y=4\times2$, $L_x\times L_y=3\times 3, L_x\times L_y=4\times 3$. Parameters $J/U = 0.1$ and (a) $T/U = 10$, (b) $T/U = 5$ and (c) $T/U = 2$. }
\label{fig:Figfse}
\end{figure}

While lattice sizes that we consider are too small to quantitatively predict  results in the thermodynamic limit, we expect that the main features of the optical conductivity that we observe remain valid. For example, at high temperature we do expect DC conductivity to take the from $\sigma_{DC}\sim J/T$, yet the exact value of the tunneling $J$ and temperature $T$ where this behavior changes into a more-complex dependence is possibly system-size dependent.  In Fig.~\ref{fig:Figfse} we compare the current-current correlation functions for three different lattice sizes at several values of temperatures. Overall, the results show good agreement and the main discrepancies are found close to $\omega\rightarrow0$.

\subsection{The grand-canonical ensemble}
\label{sec:gce}
\begin{figure}[!t]
\includegraphics[width=\linewidth]{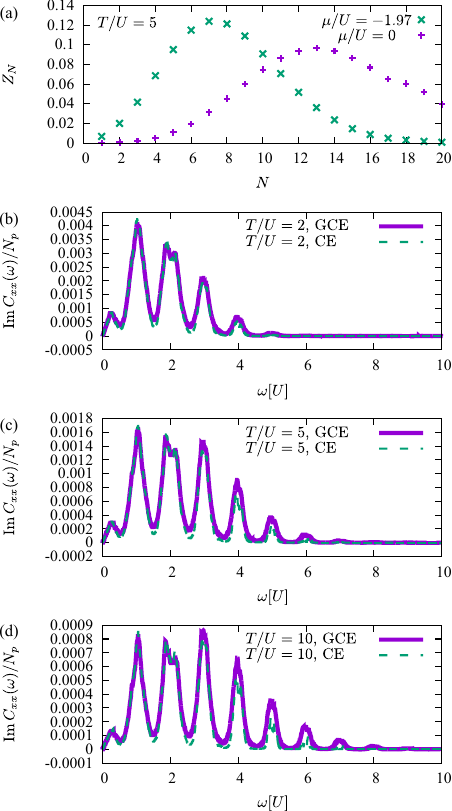}
\caption{(a) The partition function in different particle sectors, $Z_N = \mathrm{Tr} \exp(-\beta (H-\mu N))$. The chemical potential is set by the requirement $\langle N \rangle/ (L_x \times L_y) = 1.$ Imaginary part of the current-current correlation function for $J/U = 0.04$ at (b) $T/U = 2$, (c) $T/U = 5$, (d) $T/U = 10$. Lattice size $L_x \times L_y = 4\times 2$, no twisted boundary conditions.}
\label{fig:Figgce}
\end{figure}
Here we consider the grand-canonical ensemble and introduce chemical potential $\mu$
\begin{equation}
 H_{\mu} = H -\mu N,
\end{equation}
where $N$ is number of particles.
The value of the chemical potential $\mu$ is set as usual, by requiring certain filling. In Fig.~\ref{fig:Figgce} we make a comparison of results obtained in this way with the results presented in the main text within the fixed particle-number sector. We consider small lattice $L_x\times L_y = 4 \times 2$ and  up to 20 particles. We find that there are some quantitative differences, while the main qualitative features of the correlation function remain unchanged. The contribution of conductivity peaks found at $\omega = n U$ is stronger at high temperatures when we take into account particle-number fluctuations within the grand-canonical ensemble.

 \newpage
\subsection{Comparison with bubble-diagram approximation}
\label{sec:bubble}

Here we use the bubble-diagram approximation, that has been extensively used for the calculation of the conductivity within the Fermi-Hubbard model \cite{Limelette2003, Terletska2011, Vucicevic2013, Deng2013, Vucicevic2015, Vucicevic2019, Vranic2020, VucicevicPRB2021, VucicevicPRL2021, Vucicevic2023}. We find that for small lattices the bubble approximation works well only at high frequencies.
Similar relation between the full result and the bubble approximation has been observed in the fermionic Hubbard model\cite{Vucicevic2019} and more recently even in the context of Holstein model\cite{JankovicArxiv2024}.  By contrast, however, here vertex corrections appear to reduce conductivity, rather than increase it.

Within the grand-canonical ensemble we first calculate the single-particle Green's function
\begin{widetext}
\begin{equation}
 G_{\mathbf{k}}(\omega+i\delta)=\frac{1}{\mathrm{Tr} e^{-\beta (H-\mu N)}}\sum_{m, n}\frac{e^{-\beta (E_n-\mu N_n)}-e^{-\beta (E_m-\mu N_m)}}{\omega+i\delta + E_n - E_m+\mu}\langle n|b_{\mathbf{k}}|m\rangle\langle m|b_{\mathbf{k}}^{\dagger}|n\rangle,
 \label{eq:g}
\end{equation}
\end{widetext}
where $b_{\mathbf{k}}=\frac{1}{\sqrt{L_x L_y}}\sum_{\mathbf{r}}e^{i \mathbf{k} \mathbf{r}} b_{\mathbf{r}}$, and ${\mathbf{r}}$ are vectors labeling lattice sites. The eigenenergies $E_n$ and $E_m$ and eigenstates $|n\rangle$ and $|m\rangle$ are obtained in two different particle sectors with $N_n$ and $N_m = N_n + 1$ particles. From here we obtain the spectral function:
\begin{equation}
  A(\mathbf{k}, \omega)=-\frac{1}{\pi} \text{Im} G_{\mathbf{k}}(\omega+i\delta).
\end{equation}
By using the Wick theorem we obtain the bubble-diagram approximation for conductivity:
\begin{widetext}
\begin{equation}
  \text{Re}\, \sigma_{xx}(\omega) = \pi \frac{1}{L_x L_y}\sum_{\mathbf{k}}\int \, d \omega_1 \, \left(\partial_{k_x}\varepsilon_{\mathbf{k}}\right)^2  A(\mathbf{k}, \omega_1)A(\mathbf{k}, \omega_1+\omega) \frac{n_B(\omega_1) - n_B(\omega_1 + \omega)}{\omega},
   \label{eq:doubleintegral}
\end{equation}
\end{widetext}
where $n_B(x)=1/(\exp(\beta x)-1)$ is the Bose-Einstein distribution and $\varepsilon_{\mathbf{k}}=-2J \left(\cos k_x+\cos k_y\right)$ is the non-interacting dispersion relation on a two dimensional lattice.

\begin{figure*}[!ht]
\includegraphics[width=\linewidth]{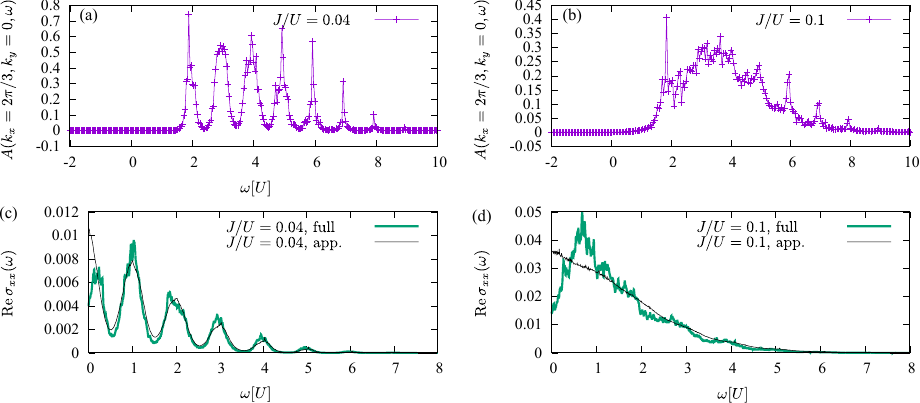}
\caption{Spectral functions for   (a)$J/U = 0.04$ and (c) $J/U = 0.1$. Optical conductivity $\mathrm{Re}\,\sigma_{xx}(\omega)$ as a function of frequency $\omega$ for (b) $J/U = 0.04$  and (d) $J/U = 0.1$. Parameters: $L_x\times L_y = 3\times 2$, $\beta U = 0.2$, $\mu/U\approx=-1.97$ such that $\langle N\rangle = 1$. }
\label{fig:FigWick}
\end{figure*}

We perform numerical test of the accuracy of this approximation by comparing results obtained from Eq.~(\ref{eq:doubleintegral}) with the full result from Eq.~(\ref{eq:sigmaxx}).  We find that for the small system sizes the approximation works well at higher frequencies $\omega$, but it fails to reproduce numerical data in the limit $\omega \rightarrow 0$ as shown in Fig.~\ref{fig:FigWick}. For $J/U = 0.04$ we find that the spectral function features separate peaks close to $\omega_n\approx-\mu+n U$, $n=0,1,2,\ldots$. From Eq.~(\ref{eq:doubleintegral}) it follows that the peaks in optical conductivity emerge when the two peaks in $A(\mathbf{k}, \omega_1)$ and $A(\mathbf{k}, \omega_1+\omega)$ overlap, as for $\omega = 0, U, 2U, \ldots$, as discussed in subsection \ref{sec:results}.

\section{Conclusion and discussion}
\label{sec:conclusions}

In this paper we investigated optical conductivity of the Bose-Hubbard model in the high-temperature regime. Based on the numerically exact calculation for small lattice sizes, we identified multiple peaks in optical conductivity stemming from the Hubbard bands at weak tunneling $J/U$. As the tunneling rate gets stronger, these peaks merge, and the conductivity takes simpler form. We analyzed the regime with resistivity linear in temperature and found that the proportionality constant is inversely proportional to the tunneling rate $J$ in the limit $J/U\rightarrow0$. Additionally, in some cases we observe two separate linear regimes of different slopes - one at lower temperature corresponding to hard-core boson behavior, and one at high temperature, corresponding to the leading order in $\beta$-expansion. Finally, we find a striking and unexpected
non-monotonic dependence of $\sigma_{\mathrm{DC}}$ on the coupling constant. At half-filling and fixed temperature $T$, above some value of $U$, $\sigma_{\mathrm{DC}}$ grows with the increasing $U$ and eventually it approaches the hard-core-boson conductivity. Further work is necessary to confirm that this feature of our results survives in the thermodynamic limit.

We expect that these results can be probed in cold-atom experiments, along the lines of Ref.~\cite{Brown2019}. For a quantitative comparison with experiments, it may turn out that additional Hamiltonian terms describing bosons in optical lattices should be taken into account in addition to the Bose-Hubbard model. Moreover, processes beyond the linear-response regime may play a role \cite{PhysRevResearch.2.043133}. In order to extend these calculations to larger system sizes, beyond-mean-field approximations \cite{Snoek_2013, He_2022, Pohl_2022, Caleffi_2020, Colussi_2022} could be considered.

Computations were performed on the PARADOX supercomputing facility (Scientific Computing Laboratory, Center for the Study of Complex Systems, Institute of Physics Belgrade). We acknowledge funding provided by the Institute of Physics Belgrade, through the grant by the Ministry of Education, Science, and Technological Development of the Republic of Serbia, as well as by the Science Fund of the Republic of Serbia, under the Key2SM project (PROMIS program, Grant No.~6066160).
J.~V. acknowledges funding by the European Research Council, grant ERC-2022-StG: 101076100.

%
\end{document}